\documentclass[11pt]{article}%
\usepackage{amsmath}
\usepackage{amsfonts}
\usepackage{amssymb}
\usepackage{graphicx}
\usepackage{fullpage}%
\providecommand{\U}[1]{\protect\rule{.1in}{.1in}}
\newtheorem{theorem}{Theorem}

\begin{document}

\title{The Kirchhoff's Matrix-Tree Theorem revisited: \\counting spanning trees with the quantum relative entropy}
\author{Vittorio Giovannetti\thanks{NEST, Scuola Normale Superiore and Istituto
Nanoscienze-CNR, Piazza dei Cavalieri 7, I-56126 Pisa, Italy;
\texttt{v.giovannetti@sns.it}.}
\and Simone Severini\thanks{Department of Physics and Astronomy, University College
London, Gower St., WC1E 6BT London, United Kingdom;
\texttt{simoseve@gmail.com}. }}
\maketitle

\begin{abstract}
By revisiting the Kirchhoff's Matrix-Tree Theorem, we give an exact formula
for the number of spanning trees of a graph in terms of the quantum relative
entropy between the maximally mixed state and another state specifically
obtained from the graph. We use properties of the quantum relative entropy to
prove tight bounds for the number of spanning trees in terms of \ basic
parameters like degrees and number of vertices.

\end{abstract}

\section{Introduction}

The \emph{Kirchhoff's Matrix-Tree Theorem} is a classic result stating that
the total number of spanning trees of a graph is exactly the determinant of
any principal minor of the Laplacian matrix of the graph. The theorem can be
traced back to 1847 \cite{ki}. The number of spanning trees of a graph, which
is also called the \emph{complexity} of the graph, is a valuable invariant
(\emph{i.e.}, a quantity that does not depend on the ordering of the
vertices). There are various and diverse applications of the number of
spanning trees: an important role in the theory of electrical networks, for
example, in computing driving point resistances \cite{bo}; a wide use in the
study of graph-theoretic problems, like routing, counting Eulerian tours,
\emph{etc. }\cite{st}; the computation of network reliability \cite{ge}; and
chemical modeling \cite{hi}. In quantum field theory, it is well-known that
the value of a Feynman integral can be written in terms of spanning trees
\cite{na}. The Tutte polynomial, object with a variety of uses in statistical
mechanics, can be characterized via spanning trees \cite{we}.

By reinterpreting the Kirchhoff's Matrix-Tree Theorem in the context of
quantum information theory, we give an exact formula to count spanning trees
based on the notion of quantum relative entropy. This function is the quantum
mechanical analog of the relative entropy. It is central in the quantification
and manipulation of quantum entanglement, quantum data compression,
communication costs, \emph{etc.} \cite{sh, ve}. It is traditionally
interpreted as a parameter to quantify the distinguishability between two
quantum states. We show that the number of spanning trees is proportional to
the distinguishability/distance between a certain density matrix associated to
the graph in context and the maximally mixed state, \emph{i.e.}, the state
with maximum von Neumann entropy, or, equivalently, maximum amount of
classical uncertainty.

By using standard machinery from quantum information theory, we study bounds
on the number of spanning trees obtained from basic quantities like the number
of vertices, edges, and degrees. We exhibit a tight bound; equality is
obtained for stars and certain multigraphs obtained by adding multiple edges
to stars. Even if the bound is tight, it performs poorly in general. A
potential improvement could be obtained by allowing different coefficients in
the quantum state associated to the graph.

The remainder of the paper is structured as follows: Section 2 contains the
mathematical setup and the main result; Section 3 gives lower and upper
bounds; Section 4 proposes a brief discussion and draws some conclusions. We
give particular attention to a plausible operational meaning for the number of
spanning trees, when considering our class of quantum states.

\section{Main result}

A \emph{graph} is an ordered pair $G=\left(  V,E\right)  $, where $V$ is a set
of elements called \emph{vertices} and $E\subseteq V\times V-\{\{i,i\}$, for
every $i\in V\}$ is set of unordered pairs of vertices called \emph{edges}.
Despite we mainly consider simple graphs, our treatment can be easily
generalized to multigraphs, \emph{i.e.}, graph with multiple edges. A graph
$H=(W,F)$ is a \emph{subgraph} of $G$ if $W\subseteq V$ and $F\subseteq E$.
The subgraph $H$ is \emph{spanning} if $W=V$. A \emph{cycle} is a graph with
set of vertices $\{0,1,...,k-1\}$ and set of edges
$\{\{i,(i+1)\operatorname{mod}k\}:i=0,1,...,k-1\}$. A \emph{tree} is a graph
without cycles as subgraphs. Denoting by $|S|$ the number of elements in a set
$S$, the \emph{degree} of a vertex $i$ is the nonnegative integer $d\left(
i\right)  =|\{j:\{i,j\}\in E\}|$. The (\emph{combinatorial}) \emph{Laplacian}
of $G$ is the $n\times n$ matrix $L=\Delta-A$, where $\Delta_{ij}=d\left(
i\right)  \delta_{ij}$ ($\delta_{ij}$ is the Kronecker delta); $A_{ij}=1$ if
$\{i,j\}\in E$ and $A_{ij}=0$, otherwise. The determinant of the $n-1\times
n-1$ matrix $L_{\ell}^{\prime}$ obtained by deleting the $\ell$-th row and
column of $L$ is independent of $\ell$. The \emph{Kirchhoff's Matrix-Tree
Theorem} (see, \emph{e.g.}, \cite{tu}, Theorem VI.29) states that for the
number of spanning trees of a graph $G$, $\tau(G)$, we have
\begin{equation}
\tau(G)=\det\left(  L_{\ell}^{\prime}\right)  . \label{kirk}%
\end{equation}

Let $\mathcal{H}$ be an Hilbert space of dimension $n$ with standard basis
$\{|1\rangle,...,|n\rangle\}$, where the vector $|i\rangle$ is associated to
the vertex $i\in V$. The \emph{volume} of $G$ is denoted and defined as
vol$\left(  G\right)  =\sum_{i\in V}d\left(  i\right)  $. By the Handshake
Lemma, vol$\left(  G\right)  =2|E|$. Denoting by $\langle i|$ the functional
that sends $|j\rangle$ to the inner product $\langle i|j\rangle$, for each
$i,j=1,...,n$, we define the matrix%
\begin{equation}
\rho=\frac{1}{\text{vol}\left(  G\right)  }\sum_{{\{i,j\}}\in E}%
|i-j\rangle\langle i-j|, \label{lap}%
\end{equation}
where $|i-j\rangle:=|i\rangle-|j\rangle$. It is promptly verified that
$\rho=L/$Tr$\left(  \Delta\right)  =L/$vol$\left(  G\right)  $ \cite{bgs}. In
what follows, we write $d=$ vol$\left(  G\right)  $, in order to simplify the
notation. We can treat $\rho$ as the state of a quantum system with assigned
Hilbert space $\mathcal{H}$, given that $\rho$ is a density matrix, being
positive-semidefinite, and trace-one. In particular, we shall use two notions
from the tool-box of quantum information theory: CPTP maps and the quantum
relative entropy.

Axiomatically, the evolution of a state $\sigma$ may be governed by a
\emph{completely positive trace preserving }(for short, \emph{CPTP}) map
$\Phi\rightarrow\Phi\left(  \sigma\right)  =\sum_{j}K_{j} \sigma K_{j}^{\dag}%
$, for some set $\{K_{j}\}$ of operators on $\mathcal{H}$ such that $\sum
_{j}K_{j}^{\dagger}K_{j}=I$, where $I$ is the identity matrix (see,
\emph{e.g.}, \cite{pa}). The \emph{von Neumann entropy} of a density matrix
$\sigma$ is defined by $S\left(  \sigma\right)  =-$Tr$\left(  \sigma\ln
\sigma\right)  $. Given density matrices $\sigma_{1}$ and $\sigma_{2}$, the
\emph{quantum} \emph{relative entropy} of $\sigma_{1}$ with respect to
$\sigma_{2}$ measures the difficulty of distinguishing between these states
and it is defined by
\[
S(\sigma_{1}\Vert\sigma_{2})=\text{Tr}\left(  \sigma_{1}\ln\sigma_{1}\right)
-\text{Tr}\left(  \sigma_{1}\ln\sigma_{2}\right)  .
\]

The Kirchhoff's Matrix-Tree Theorem tells us that the number of spanning trees
of a graph $G\,$on $n$ vertices, $\tau(G)$, can be expressed in terms of the
density matrix $\rho$. From Eqs.~(\ref{kirk}) and (\ref{lap}), we are prompted
to the next statement:

\begin{theorem}
Let $G$ be a graph on $n$ vertices and $d/2$ edges. The number of spanning
trees of $G$ is%
\begin{equation}
\tau(G)=\left(  \tfrac{d}{n-1}\right)  ^{n-1}e^{-(n-1)S(\frac{\Pi_{\ell}}%
{n-1}\Vert\Phi_{\ell}(\rho))}, \label{ggf11}%
\end{equation}
and also
\begin{equation}
\tau(G)=\left(  \tfrac{d}{n-1}\right)  ^{n-1}e^{-\tfrac{n-1}{n}\sum_{\ell
=1}^{n}S(\frac{\Pi_{\ell}}{n-1}\Vert\Phi_{\ell}(\rho))}. \label{ggf110}%
\end{equation}
Here, $\Pi_{\ell}$ is a projection operator onto the $n-1$ dimensional space
spanned by all the elements of the standard basis associated to $G$, but the
$\ell$-th; $\Phi_{\ell}$ is the CPTP map defined by
\begin{equation}
\Phi_{\ell}(\rho)=\Pi_{\ell}\rho\Pi_{\ell}+Q_{\ell}\rho Q_{\ell}, \label{phil}%
\end{equation}
where
\[
Q_{\ell}=I-\Pi_{\ell}=|\ell\rangle\langle\ell|\;,
\]
is the complementary projector of $\Pi_{\ell}$.
\end{theorem}

(In matrix theory, the above transformation is called \emph{pinching} and
consists of removing all the off-diagonal terms related to the $\ell$-th
vertex of the graph.) The derivation of Eq.~(\ref{ggf11}) is simple and
ultimately follows from the identity
\[
\exp\left(  -S(I/n\Vert M)\right)  =N(\det M)^{1/n},
\]
which holds for all positive semidefinite $n\times n$ matrices $M$. For a more
explicit derivation, we firstly construct the $n\times n$ matrix which has
zeros in the $\ell$-th row and in the $\ell$-th column, but it is identical to
the Laplacian $L$ in the remaining entries. We can express it as
\begin{equation}
\Pi_{\ell}L\Pi_{\ell}=d\Pi_{\ell}\rho\Pi_{\ell}. \label{philL}%
\end{equation}
By construction, the matrix $\Pi_{\ell}L\Pi_{\ell}$ will have the same
spectrum of $L_{\ell}^{\prime}$ \emph{plus} an extra zero eigenvalue. Recall
in fact that $L_{\ell}^{\prime}$ is an $(n-1)\times(n-1)$ matrix, while
$\Pi_{\ell}L\Pi_{\ell}$ is $n\times n$. Such an extra zero eigenvalue is
associated with the eigenvector $|\ell\rangle$. Therefore, denoting by
$\lambda_{i}^{\prime}$ the eigenvalues of $L_{\ell}^{\prime}$, we can write
\begin{equation}
\tau(G)=\det\left(  L_{\ell}^{\prime}\right)  =\prod_{i\in\mathcal{S}^{\prime
}}\lambda_{i}^{\prime}. \label{gg44}%
\end{equation}
where the set $S^{\prime}$ contains all eigenvalues of $\Pi_{\ell}L\Pi_{\ell}$
but the extra zero. This formula implies%
\[
\ln\tau(G)=\sum_{i\in\mathcal{S}^{\prime}}\ln\lambda_{i}^{\prime}%
=\text{Tr}(\Pi_{\ell}\ln(\Pi_{\ell}L\Pi_{\ell})).
\]
(We adopt the standard convention $0\ln0=0$.) Thus, by Eqs.~(\ref{phil}%
,\ref{philL}),
\begin{align*}
\ln\tau(G)  &  =({n-1})\text{Tr}(\tfrac{\Pi_{\ell}}{n-1}\ln(d\Pi_{\ell}\rho
\Pi_{\ell}))\\
&  =({n-1})\text{Tr}(\tfrac{\Pi_{\ell}}{n-1}\ln(\Pi_{\ell}\rho\Pi_{\ell
}))+(n-1)\ln d\\
&  =({n-1})\text{Tr}(\tfrac{\Pi_{\ell}}{n-1}\ln\Phi_{\ell}(\rho))+(n-1)\ln d,
\end{align*}
where in the last identity we used the fact that since $Q_{\ell}$ and
$\Pi_{\ell}$ are orthogonal projectors we have
\[
\ln(\Pi_{\ell}\rho\Pi_{\ell}+Q_{\ell}\rho Q_{\ell})=\ln(\Pi_{\ell}\rho
\Pi_{\ell})\Pi_{\ell}+\ln(Q_{\ell}\rho Q_{\ell})Q_{\ell}.
\]
Noticing that $\Pi_{\ell}/(n-1)$ is positive semi-definite and has trace one,
we can identify this operator with a density matrix. Indeed, it is the density
matrix describing a maximally mixed state of the subspace orthogonal to
$|\ell\rangle$. It follows that
\begin{align*}
\text{Tr}(\tfrac{\Pi_{\ell}}{n-1}\ln\Phi_{\ell}(\rho))  &  =-S(\tfrac
{\Pi_{\ell}}{n-1})-S(\tfrac{\Pi_{\ell}}{n-1}\Vert\Phi_{\ell}(\rho))\\
&  =-\ln(n-1)-S(\tfrac{\Pi_{\ell}}{n-1}\Vert\Phi_{\ell}(\rho)).
\end{align*}
Hence,
\[
\ln\tau(G)=-({n-1})S(\tfrac{\Pi_{\ell}}{n-1}\Vert\Phi_{\ell}(\rho
))\;+(n-1)\ln\tfrac{\text{vol}\left(  G\right)  }{n-1},
\]
which reduces to Eq.~(\ref{ggf11}) by exponentiating it -- Eq.~(\ref{ggf110})
is then a trivial consequence of Eq.~(\ref{ggf11}).

If the graph does not have isolated vertices (\emph{i.e.} if $\Delta$ is
invertible, or $d\left(  i\right)  >0$ for every $i\in V$) an alternative (but
equivalent) expression can be obtained as follows:
\begin{align}
\tau(G)  &  =\left(  \tfrac{d}{n}\right)  ^{n}\frac{e^{-nS(\tfrac{I}{n}%
\Vert\Phi_{\ell}(\rho))}}{\Delta_{\ell}}\label{uno}\\
&  =\left(  \tfrac{d}{n}\right)  ^{n}\frac{e^{-\sum_{\ell=1}^{n}S(\tfrac{I}%
{n}\Vert\Phi_{\ell}(\rho))}}{\det({\Delta})^{1/n}}, \label{due}%
\end{align}
where, for $\ell\in\{1,\cdots,n\}$, $\Delta_{\ell}$ is the $\ell$-th diagonal
element of the degree matrix $\Delta$. If $\Delta$ is not invertible,
\emph{i.e.}, if it has at least a zero diagonal entry, the graph $G$ has no
spanning tree. In this case, one can verify that Eq.~(\ref{uno}) still applies
under the assumption that $\ell$ is not an isolated vertex of the graph. On
the other hand, Eq.~(\ref{due}) cannot be used, as the denominator diverges.
To include this case, we should simply say that if $\det\Delta=0$ then
$\tau(G)=0$, while if $\det\Delta\neq0$ then Eq.~(\ref{due}) applies. To prove
the above identity, it is sufficient to verify Eq.~(\ref{uno}), since
Eq.~(\ref{due}) is obtained by multiplying such term over all possible values
of $\ell$ and by taking $n$-th root of the result. For Eq.~(\ref{uno}), we go
back to Eq.~(\ref{gg44}) and notice that
\begin{align}
\ln\tau(G)  &  =\text{Tr}(\Pi_{\ell}\ln(d\Pi_{\ell}\rho\Pi_{\ell}))\nonumber\\
&  =\text{Tr}({\Pi_{\ell}}\ln(\Pi_{\ell}\rho\Pi_{\ell}))+(n-1)\ln d\nonumber\\
&  =\text{Tr}(\ln\Phi_{\ell}(\rho))-\ln(\Delta_{\ell}/d)+(n-1)\ln d\nonumber\\
&  =\text{Tr}(\ln\Phi_{\ell}(\rho))-\ln(\Delta_{\ell})+n\ln d,
\end{align}
where we used the fact that
\begin{align*}
\ln\Phi_{\ell}(\rho)  &  =\Pi_{\ell}\ln(\Pi_{\ell}\rho\Pi_{\ell})+Q_{\ell}%
\ln(Q_{\ell}\rho Q_{\ell})\\
&  =\Pi_{\ell}\ln(\Pi_{\ell}\rho\Pi_{\ell})+Q_{\ell}\ln(\Delta_{\ell}/d).
\end{align*}
Eq.~(\ref{uno}) finally follows from
\[
\text{Tr}(\ln\Phi_{\ell}(\rho))=n\text{Tr}(\tfrac{I}{n}\ln\Phi_{\ell}%
(\rho))=-nS(\tfrac{I}{n}\Vert\Phi_{\ell}(\rho))-n\ln n.
\]
Eqs.~(\ref{ggf11},\ref{ggf110},\ref{uno},\ref{due}) allow us to express
$\tau(G)$ in terms of the density matrix $\rho$ and the relative entropy. It
is worth noticing that it is possible to recast these equations in the
following form:
\[
\tau(G)=\left(  \tfrac{d-\Delta_{\ell}}{n-1}\right)  ^{n-1}e^{-(n-1)S(\frac
{\Pi_{\ell}}{n-1}\Vert\rho_{\ell}^{\prime})};
\]
now $\rho_{\ell}^{\prime}$ is the density matrix obtained by normalizing
$L_{\ell}^{\prime}$, \emph{i.e.},
\[
\rho_{\ell}^{\prime}=\frac{L_{\ell}^{\prime}}{d-\Delta_{\ell}}\sim\frac
{\Pi_{\ell}L\Pi_{\ell}}{d-\Delta_{\ell}},
\]
where $\sim$ denotes equivalence. (Notice that the last term is an $n\times n$
matrix with a zero column and a zero row.) The derivation of this expression
is exactly on the same line as the previous ones.

\section{Bounds}

We can derive lower and upper bounds on $\tau(G)$ by exploiting known facts
about the quantum relative entropy. On the other hand, deriving meaningful,
simple lower bounds on $\tau(G)$ with our technique does not seem to be
natural. A first attempt based on the monotonicity of the relative entropy
fails to produce even a trivial result, which is $\tau(G)\geqslant0$. We
discuss this case here only as an exercise. Firstly, observe that \emph{i)
}$\Phi_{\ell}$ is a unital quantum channel (\emph{i.e.}, it maps the identity
operator into itself $\Phi_{\ell}(I)=I$) and that \emph{ii)} the relative
entropy is not increasing under CPTP maps, an important and nontrivial
property:
\[
S(\rho_{1}\Vert\rho_{2})\geqslant S(\Phi(\rho_{1})\Vert\Phi(\rho_{2})).
\]
Using these two facts we can exhibit a lower bound for $\tau(G)$:
\begin{equation}
S(\tfrac{I}{n}\Vert\Phi_{\ell}(\rho))=S(\Phi_{\ell}(\tfrac{I}{n})\Vert
\Phi_{\ell}(\rho))\leqslant S(\tfrac{I}{n}\Vert\rho)=-\ln n-\tfrac{1}{n}%
\sum_{j=1}^{n}\ln\lambda, \label{hh}%
\end{equation}
where $\lambda_{j}$ is the $j$-th eigenvalues of $\rho$. Replacing this into
Eq.~(\ref{uno}), we thus get
\begin{align}
\tau(G)  &  =\frac{d^{n}}{\Delta_{\ell}n^{n}}e^{-nS(\tfrac{I}{n}\Vert
\Phi_{\ell}(\rho))}\geqslant\frac{d^{n}}{\Delta_{\ell}}e^{\sum_{j=1}^{n}%
\ln\lambda_{j}}\nonumber\\
&  =\frac{d^{n}\det(\rho)}{\Delta_{\ell}}=\frac{\det(L)}{\Delta_{\ell}},
\label{uno1}%
\end{align}
which is true for all $\ell$. Therefore taking the minimum over $\ell$, we can
write
\begin{equation}
\tau(G)\geqslant\frac{\det(L)}{\delta_{min}}, \label{uno2}%
\end{equation}
where $\delta_{min}$ denotes the minimum degree of the graph. Similarly,
replacing Eq.~(\ref{hh}) into Eq.~(\ref{due}), we get the inequality
$\tau(G)\geqslant\det(L)/(\det\Delta)^{1/n}$. Notice that these lower bounds
are both trivial. Indeed, since $L$ has at least a null eigenvalue we have
that $\det(L)=0$ and thus the above inequalities simply states that
$\tau(G)\geqslant0$. It is worth observing that one could in principle get
something less trivial by replacing the density $\rho$ which appears in the
\emph{r.h.s.} of Eq.~(\ref{hh}) with a generic $\bar{\rho}$ state which
satisfies the condition $\Phi_{\ell}(\bar{\rho})=\Phi_{\ell}(\rho)$ (notice
that such $\bar{\rho}$ can be easily constructed and that they have the
property of having the same diagonal elements of $\rho$). This would allow us
to rewrite Eq.~(\ref{uno2}) as $\tau(G)\geqslant\det(\bar{L})/\delta_{min}$,
where $\bar{L}=\bar{\rho}d$ (in this case $\det(\bar{L})$ needs not to be zero).

Using the results in \cite{he}, we can determine upper bound on $S(\cdot
\Vert\cdot)$ which, through our expressions, will result on bounds on
$\tau(G)$. None of the bounds seem to be particularly relevant (they are too
weak and require to compute quantities which are not easily computable). As an
example we just report here one of them which is obtained through the
inequality
\[
S(\rho_{1}\Vert\rho_{2})\leqslant\frac{\text{Tr}((\rho_{1}-\rho_{2})^{2}%
)}{\lambda_{min}(\rho_{2})},
\]
where $\lambda_{min}(\rho_{2})$ is the smallest nonzero eigenvalues of
$\rho_{2}$. We apply this inequality to the identity of Eq.~(\ref{uno}). In
this case, $\rho_{1}=I/n$ and $\rho_{2}=\Phi_{\ell}(\rho)$. Hence, exploiting
the properties of $\rho$,
\[
\text{Tr}((I/n-\Phi_{\ell}(\rho))^{2})=-1/n+\frac{\text{Tr}(\Delta
^{2})+\text{Tr}(\Delta)-2\Delta_{\ell}}{d^{2}}\;.
\]
The negative fact is that we end up to compute the smallest eigenvalue of
$\Phi_{\ell}(\rho)$ and this is only slightly less time demanding than
computing the determinant (and of course the determinant of $\Phi_{\ell}%
(\rho)$ is simply $\Delta_{\ell}\tau(G)$).

In the following, we derive several upper bounds for $\tau(G)$. Before
entering into the details of the derivation, we summarize the most relevant bounds:

\begin{theorem}
Let $G$ be a graph on $n$ vertices and maximum degree $\Delta_{\max}$. Then%
\[
\tau(G)\leqslant\min\{\tau_{0},\tau_{A},\cdots,\tau_{E}\},
\]
where%
\[%
\begin{array}
[c]{lll}%
\tau_{0}=\frac{\det(\Delta)}{\Delta_{\max}}, & \qquad & \tau_{A}={\det
(\Delta)}^{1-1/n},\\
\tau_{B}=\text{\emph{Tr}}(\Delta/n)^{n-1}, & \qquad & \tau_{C}=\frac
{\text{\emph{Tr}}(\Delta/n)^{n}}{\det({\Delta)}^{1/n}},\\
\tau_{D}=\frac{\text{\emph{Tr}}(\Delta/n)^{n}}{\Delta_{\max}}, & \qquad &
\tau_{E}=\left(  \frac{\text{\emph{Tr}}(\Delta)-\Delta_{\max}}{n-1}\right)
^{n-1}.
\end{array}
\]
{The bound $\tau_{0}$ i}s tight in{ the sense that for each value of $\tau
_{0}$ we can construct a graph }$H$ {with an arbitrary number of vertices
$n\geqslant2$ such that }%
\[
\tau(H)={\tau_{0}\left(  H\right)  .}%
\]

\end{theorem}

As a remark, it is worth observing that we can verify the following
inequalities hold:%
\[%
\begin{tabular}
[c]{l}%
$\tau_{0}\leqslant\tau_{A}\leqslant\tau_{B}\leqslant\tau_{C},$\\
$\tau_{0}\leqslant\tau_{D}\leqslant\tau_{B},$\\
$\tau_{0}\leqslant\tau_{E}\leqslant\tau_{B}.$%
\end{tabular}
\]
Also, there is not a definite ordering between $\tau_{D}$ and $\tau_{A}$. A
further upper bound, worse than $\tau_{A}$, is
\[
\tau_{F}=\left(  \tfrac{n}{n-1}\right)  ^{n-1}\det(\Delta)^{1-1/n}.
\]
All the bounds in the theorem should be compared with the trivial value
$\tau_{trivial}=\left(
\begin{array}
[c]{c}%
\text{Tr}(\Delta/2)\\
n-1
\end{array}
\right)  $, which simply follows by observing that any spanning tree has
exactly $n-1$ edges and that Tr$(\Delta/2)$ is the total number of edges of
the graph. It turns out that there is not a definitive ordering among
$\tau_{trivial}$ and the bounds that we derived above. Indeed, our bounds
perform better on some graphs only.

As an example, let us consider the following cases:

\begin{itemize}
\item The complete graph, $K_{n}$. According to the Cayley formula,
$\tau(K_{n})=n^{n-2}$. For $K_{n}$, we have $\Delta=(n-1)I$, $\Delta_{\max
}=n-1$, Tr$(\Delta)=n(n-1)$, and $\det\Delta=(n-1)^{n}$. Hence, all our bounds
coincide, \emph{i.e.}, $\tau_{0}=\tau_{A}=\tau_{B}=\tau_{C}=\tau_{D}=\tau
_{E}=(n-1)^{n-1}$, while $\tau_{trivial}=\left(
\begin{array}
[c]{c}%
n(n-1)/2\\
n-1
\end{array}
\right)  $, which for $n\geqslant7$ is already larger than $(n-1)^{n-1}$.

\item The star, $K_{1,n-1}$. In this case, $\Delta$ has one eigenvalue equal
to $n-1$ (in fact, $\Delta_{max}=n-1$) and $n-1$ eigenvalues equal to $1$.
Clearly, $\tau(K_{1,n-1})=1$. We have that $\tau_{0}=\tau_{E}=1$, a tight
bound, while $\tau_{A}=(n-1)^{n-1}$, $\tau_{B}=\left(  \tfrac{2(n-1)}%
{n}\right)  ^{n-1}$, $\tau_{C}=\left(  \tfrac{2(n-1)}{n}\right)  ^{n}\tfrac
{1}{(n-1)^{1/n}}$, $\tau_{D}=\left(  \tfrac{2(n-1)}{n}\right)  ^{n}\tfrac
{1}{n-1}$. Notice that in this case $\tau_{A}<\tau_{D}$ for $n\geqslant6$ and
$\tau_{A}>\tau_{D}$ for $n<6$.

\item Consider the graph with $n=4$ vertices and edges
$\{1,2\},\{2,3\},\{3,4\},\{2,4\}$. One can easily verify that $\tau(G)=3$. For
this graph, we have $\tau_{0}=\tau_{trivial}=4$, while $\tau_{A}\simeq6.45$,
$\tau_{B}=8$, $\tau_{C}=8.6$, $\tau_{D}\simeq5.33$, and $\tau_{E}\simeq4.62$.
Notice that in this case $\tau_{A}>\tau_{D}$.
\end{itemize}

To prove the optimality of ${\tau_{0}}$, we firstly observe that each graph
$G$ with $n=2$ and multiple edges saturate the bound. Those graphs are only
characterized by the number $k$ of edges which connects the two element of
$V$: therefore $\Delta=$ diag$(k,k)$ and $\tau_{0}(G)=k$. For multigraphs on
an arbitrary number of vertices $n>2$, we consider stars with multiple edges
only between an arbitrary but fixed pair of adjacent vertices. \emph{W.l.o.g.}
we may assume $k$ edges between vertices $1$ and $2$ only. These multigraphs
are denoted by $K_{1,n,k}$. It is clear that $\tau(K_{1,n,k})=k$. Furthermore,
since $\Delta=$ diag$(k+(n-1),k,1,\cdots,1)$, then also $\tau_{0}%
(K_{n,1,k})=k$.

Both these bounds follow from the Klein inequality: the relative entropy of
two states is always positive semi-definite. On the basis of Eq.~(\ref{uno}),
we can conclude that
\begin{equation}
\tau(G)\leqslant\frac{d^{n}}{n^{n}\Delta_{\ell}}=\frac{\text{Tr}(\frac{\Delta
}{n})^{n}}{\Delta_{\ell}}, \label{duedueprimo}%
\end{equation}
for all $\ell$. Minimizing the \emph{r.h.s.} with respect to such index, we
can then write
\begin{equation}
\tau(G)\leqslant\frac{d^{n}}{n^{n}\Delta_{\max}}=\frac{\text{Tr}(\frac{\Delta
}{n})^{n}}{\Delta_{\max}}=\tau_{D}, \label{duedueprimomax}%
\end{equation}
Similarly, from Eq.~(\ref{due}),
\begin{equation}
\tau(G)\leqslant\frac{d^{n}}{n^{n}\det({\Delta})^{1/n}}=\frac{\text{Tr}%
(\frac{\Delta}{n})^{n}}{\det({\Delta})^{1/n}}=\tau_{C}. \label{duedue}%
\end{equation}
Since $\Delta_{\max}\geqslant\det({\Delta)}^{1/n}\geqslant\delta_{\min}$, if
follows that $\tau_{D}\leqslant\tau_{C}$.

A more refined bound can be obtained by exploiting the following inequality
(see \cite{ni}):%
\[
\sum_{i}{p_{i}}S(r_{i}\Vert s_{i})\geqslant S(\sum_{i}p_{i}r_{i}\Vert\sum
_{j}q_{j}s_{j})-\sum_{i}p_{i}\ln(p_{i}/q_{i});
\]
this is valid if $r_{i},s_{i}$ are density matrices and $p_{i}$ and $q_{i}$
are generic probability distributions. Let us apply this to $\sum_{\ell=1}%
^{n}S(I/n\Vert\Phi_{\ell}(\rho))$, with $p_{i}=q_{i}=1/n$:
\[
\frac{1}{n}\sum_{\ell}S(\tfrac{I}{n}\Vert\Phi_{\ell}(\rho))\geqslant
S(\tfrac{I}{n}\Vert\sum_{\ell}\Phi_{\ell}(\rho)/n)-(1/n)\ln(n/n)=S(\tfrac
{I}{n}\Vert\sigma).
\]
where
\begin{align}
\sigma &  =\sum_{\ell}\Phi_{\ell}(\rho)/n=\sum_{\ell}(\Pi_{\ell}\rho\Pi_{\ell
}+Q_{\ell}\rho Q_{\ell})/n=\frac{\Delta}{d}-\frac{n-2}{n}\frac{A}%
{d}\label{sigma}\\
&  =\frac{L}{d}+\frac{2}{n}\frac{A}{d}.\nonumber
\end{align}
With the use of Eq.~(\ref{due}), we finally can write
\begin{align}
\tau(G)  &  \leqslant\frac{d^{n}}{n^{n}\det({\Delta})^{1/n}}e^{-nS(\tfrac
{I}{n}\Vert\sigma)}=\frac{d\det(\sigma)}{\det({\Delta})^{1/n}}\nonumber\\
&  =\frac{\;\det(\Delta-\tfrac{n-2}{n}A)}{\det({\Delta})^{1/n}}=\frac
{\;\det(L+2A/{n})}{\det({\Delta}^{1)/n}}. \label{due221}%
\end{align}

The bound is interesting but it still involves the computation of a
determinant. We can however do better by using again the fact that the
relative entropy is decreasing under the action of CPTP maps. Consider the
CPTP map
\begin{equation}
\Psi=\frac{1}{n}\sum_{\ell=1}^{n}\Phi_{\ell}; \label{defpsi}%
\end{equation}
\[
\Psi(\Delta)=\Delta,\qquad\Psi(A)=\frac{n-2}{n}A.
\]
Now,
\[
\Psi^{k}(\sigma)=\Psi^{k+1}(\rho)=\frac{\Delta}{d}-\left(  \frac{n-2}%
{n}\right)  ^{k+1}\frac{A}{d},
\]
where $\Psi^{k}$ represent the CPTP map obtained by concatenating $k$ times
the map $\Psi$ (\emph{i.e.} $\Psi$ is applied to $\sigma$ exactly $k$ times).
For $k\gg1$ this yields (in any norm),
\[
\lim_{k\rightarrow\infty}\Psi^{k}(\sigma)=\frac{\Delta}{d}.
\]
Therefore,
\[
S(\tfrac{I}{n}\Vert\sigma)\geqslant S(\Psi^{k}(\tfrac{I}{n})\Vert\Psi
^{k}(\sigma))=S(\tfrac{I}{n}\Vert\Psi^{k}(\sigma))=S(\tfrac{I}{n}\Vert
\frac{\Delta}{d}-\left(  \frac{n-2}{n}\right)  ^{k+1}\frac{A}{d}),
\]
which in the limit of large $k$ gives
\[
S(\tfrac{I}{n}\Vert\sigma)\geqslant S(\tfrac{I}{n}\Vert\frac{\Delta}{d}).
\]
Replacing this into the first line of Eq.~(\ref{due221}), we finally obtain
\begin{align}
\tau(G)  &  \leqslant\frac{d^{n}}{n^{n}\det({\Delta})^{1/n}}e^{-nS(\tfrac
{I}{n}\Vert\sigma)}\nonumber\\
&  \leqslant\frac{d^{n}}{n^{n}\det({\Delta})^{1/n}}e^{-nS(\tfrac{I}{n}%
\Vert\frac{\Delta}{d})}=\det({\Delta})^{1-1/n}=\tau_{A}. \label{due222444}%
\end{align}
This upper bound, as the one of Eq.~(\ref{duedue}), is just a function of the
matrix $\Delta$. It is natural to ask which of the two bounds is better: the
bound (\ref{due222444}) is always better than (\ref{duedue}). To prove this,
take
\begin{equation}
\frac{\tau_{A}}{\tau_{C}}=\frac{\det\Delta}{\text{Tr}(\tfrac{\Delta}{n})^{n}%
}\leqslant1, \label{ineqp}%
\end{equation}
which can be easily verified by exploiting the fact that $\ln(x)$ is a concave
function of its argument. This clarify the relation between $\tau_{A}$ and
$\tau_{C}$.

For the bound $\tau_{B}$, we start form observing that Eq.~(\ref{uno}) and
(\ref{due}) imply that the following inequality should apply for any $\ell$:
\[
\Delta_{\ell}=\det(\Delta)^{1/n}\frac{e^{-nS(\tfrac{I}{n}\Vert\Phi_{\ell}%
(\rho))}}{e^{-\sum_{\ell=1}^{n}S(\tfrac{I}{n}\Vert\Phi_{\ell}(\rho))}%
}\leqslant\frac{\det(\Delta)^{1/n}}{e^{-\sum_{\ell=1}^{n}S(\tfrac{I}{n}%
\Vert\Phi_{\ell}(\rho))}}.
\]
where the last step is (again) a consequence of the Klein inequality. Now,
\[
\text{Tr}(\Delta)\leqslant\frac{n\det(\Delta)^{1/n}}{e^{-\sum_{\ell=1}%
^{n}S(\tfrac{I}{n}\Vert\Phi_{\ell}(\rho))}}\Longrightarrow\frac{e^{-\sum
_{\ell=1}^{n}S(\tfrac{I}{n}\Vert\Phi_{\ell}(\rho))}}{\det(\Delta)^{1/n}%
}\leqslant\frac{n}{\text{Tr}(\Delta)}.
\]
Recalling that Tr$(\Delta)=$ vol$\left(  G\right)  $ and replacing this into
Eq.~(\ref{due}), we have
\[
\tau(G)\leqslant\text{Tr}(\tfrac{\Delta}{n})^{n-1}=\tau_{B}.
\]
We can now compare $\tau_{B}$ to $\tau_{A}$ and $\tau_{C}$. Begin by writing
\[
\frac{\tau_{B}}{\tau_{C}}=\left(  \frac{\det(\Delta)}{\text{Tr}(\tfrac{\Delta
}{n})^{n}}\right)  ^{1/n}=\left(  \frac{\tau_{A}}{\tau_{C}}\right)  ^{1/n}.
\]
Thanks to Eq.~(\ref{ineqp}), we can thus conclude that
\[
1\geqslant\frac{\tau_{B}}{\tau_{C}}\geqslant\frac{\tau_{A}}{\tau_{C}},
\]
which implies $\tau_{C}\geqslant\tau_{B}\geqslant\tau_{A}$.

For $\tau_{0}$, we start from Eq.~(\ref{uno}) which we specify for
$\Delta_{\ell}=\Delta_{max}$, \emph{i.e.},
\begin{equation}
\tau(G)=\left(  \frac{d}{n}\right)  ^{n}\frac{e^{-nS(\tfrac{I}{n}\Vert
\Phi_{\hat{\ell}}(\rho))}}{\Delta_{\max}}, \label{bound01}%
\end{equation}
where $\hat{\ell}$ is the value of $\ell$ achieving $\Delta_{\max}$. We then
use the monotonicity of the relative entropy to write
\begin{align*}
S(\tfrac{I}{n}\Vert\Phi_{\hat{\ell}}(\rho))  &  \geqslant S(\Psi^{k}(\tfrac
{I}{n})\Vert\Psi^{k}(\Phi_{\hat{\ell}}(\rho)))=S(\tfrac{I}{n}\Vert\Phi
_{\hat{\ell}}(\Psi^{k}(\rho)))\\
&  \simeq S(\tfrac{I}{n}\Vert\Phi_{\hat{\ell}}(\tfrac{\Delta}{d}))=S(\tfrac
{I}{n}\Vert\tfrac{\Delta}{d}),
\end{align*}
where $\Psi$ is the unital CPTP map introduced in Eq.~(\ref{defpsi}). In
writing the last passage of the first line we used the fact that since for
each $\ell$ and $\ell^{\prime}$ the channels $\Phi_{\ell}$ and $\Phi
_{\ell^{\prime}}$ commute, also $\Psi$ (and thus $\Psi^{k}$) commute with
$\Phi_{\ell}$. The first identity in the second line is obtained for large $k$
using the fact that $\Psi^{k}(\rho)\simeq\Delta/$vol$\left(  G\right)  $, and
finally the last passage follows from the fact that for $\ell$, $\Phi_{\ell
}(\Delta)=\Delta$. Replacing this into Eq.~(\ref{bound01}), we finally get
\[
\tau(G)\leqslant\left(  \frac{d}{n}\right)  ^{n}\frac{e^{-nS(\tfrac{I}{n}%
\Vert\tfrac{\Delta}{d})}}{\Delta_{\max}}=\left(  \frac{d}{n}\right)  ^{n}%
\frac{n^{n}\det(\tfrac{\Delta}{d})}{\Delta_{\max}}=\frac{\det(\Delta)}%
{\Delta_{\max}}=\tau_{0}.
\]
To clarify the relations with the other bounds, we need only to observe that
\[
\frac{\tau_{0}}{\tau_{A}}=\frac{\det(\Delta)^{1/n}}{\Delta_{\max}}\leqslant1.
\]
To derive $\tau_{E}$ we use Eq.~(\ref{ggf11}) and the Klein inequality, as we
have already done above. This yields
\[
\tau(G)=\left(  \tfrac{d-\Delta_{\ell}}{n-1}\right)  ^{n-1}e^{-(n-1)S(\frac
{\Pi_{\ell}}{n-1}\Vert\rho_{\ell}^{\prime})}\leqslant\left(  \tfrac
{d-\Delta_{\ell}}{n-1}\right)  ^{n-1}=\left(  \tfrac{\text{Tr}(\Delta
)-\Delta_{\ell}}{n-1}\right)  ^{n-1},
\]
for all $\ell$. The case $\Delta_{\ell}=\Delta_{\max}$ gives $\tau_{E}$;
$\tau_{E}>\tau_{0}$ follows from the concavity of $\log(x)$; $\tau_{E}%
<\tau_{B}$ follows from $\Delta_{\max}\geqslant$Tr$(\Delta)/n$.

By repeating the derivation of $\tau_{A}$, starting from Eq. (\ref{ggf11})
instead of Eq. (\ref{uno}), we can derive a new upper bound, $\tau_{F}$.
However, this turns out to be weaker than $\tau_{A}$. Specifically, we get the
following inequality:
\begin{equation}
\tau(G)\leqslant\left(  \frac{n}{n-1}\right)  ^{n-1}\det(\Delta)^{1-1/n}%
=\tau_{F}. \label{otherbound}%
\end{equation}
To verify this, we use the joint convexity of the relative entropy and then
its monotonicity under CPTP maps:
\begin{align}
\frac{1}{n}\sum_{\ell=1}^{n}S(\tfrac{\Pi_{\ell}}{n-1}\Vert\Phi_{\ell}(\rho))
&  \geqslant S(\sum_{\ell=1}^{n}\frac{\Pi_{\ell}/(n-1)}{n}\Vert\sum_{\ell
=1}^{n}\frac{\Phi_{\ell}(\rho)}{n})\nonumber\\
&  =S(I/n\Vert\Psi(\rho))\geqslant S(\Psi^{k}(I/n)\Vert\Psi^{k+1}%
(\rho))\nonumber\\
&  =S(I/n\Vert\Psi^{k+1}(\rho))\simeq S(I/n\Vert\Delta/d),\nonumber
\end{align}
where $\Psi$ is the unital CPTP map introduced in Eq.~(\ref{defpsi}) and where
the last identity holds for $k\rightarrow\infty$. Replacing this inequality
into Eq.~(\ref{ggf11}), we finally obtain
\[
\tau(G)\leqslant\left(  \frac{d}{n-1}e^{-S(I/n\Vert\Delta/d)}\right)
^{n-1}=\left(  \frac{d}{n-1}n\det(\Delta/d)^{1/n}\right)  ^{n-1},
\]
which coincides with (\ref{otherbound}).

\section{Conclusions}

We comment on the expressions that we have derived to see if these could help
us in providing an operational meaning to $\tau(G)$ or more generally to
$\rho$. We notice that all the expressions derived so far allow us to write
$\tau(G)$ as a product of two quantities: a first term which is typically
greater than one and which depends only on the degree matrix of the graph; an
exponential which is always smaller than one and which depends on the relative
entropy between a density matrix obtained via some processing of $\rho$ and a
totally mixed density operator. In particular, we may point out that, apart
from the cases of (\ref{uno}) and (\ref{due}) where there is an extra factor
inversely proportional to $\Delta_{\ell}$, the first contribution goes like
$\sim(d/n)^{n}$, where $d$ is the sum of the degrees. Since the quantity $d/n$
measures the average degree, $(d/n)^{n}$ counts the average number of
independent walks one would obtain while hopping randomly $n-1$ times along
the edges of the graph while starting from a generic vertex. This number is
clearly much larger than the number of spanning tree of $G$ (the latter being
just a proper subset of the trajectories generated by the hopping). The
exponential term of our expressions can then be interpreted as the fraction of
the trajectories which indeed correspond to a spanning tree of $G$. Their
values are obtained by computing the relative entropy between a density matrix
associated with $G$ and a totally mixed state (either $n$ dimensional or $n-1$ dimensional).

Invoking the quantum Stein's Lemma we can provide these quantities with an
operational meaning in the context of the (asymmetric) quantum hypothesis
testing problem (see \cite{QHT}). It is known that $S(\rho_{1}\Vert\rho_{2})$
represents the optimal upper bound on type-II error rate for any sequence of
measurements used to decide whether a given state is $\rho_{1}$ or $\rho_{2}$,
under the conditions of bounded type-I error. Let us recall that in the
hypothesis testing problem, type-II (or \emph{false negative}) errors are
those in which we have mistaken $\rho_{2}$ with $\rho_{1}$; for type-I
(or\ \emph{false positive}) errors are those in which we have mistaken
$\rho_{1}$ with $\rho_{2}$. This implies that the exponential quantities in
our expressions quantify how different is the quantum state associated to our
graph from the totally mixed state.

The study of the relation between relative entropy and the complexity of a
graph which we have carried on in this paper is probably not exhaustive. It is
plausible to believe that allowing freedom to add arbitrary coefficients to
the projectors in the definition of $\rho$ we can push the relative entropy
closer to the effective number of spanning trees or even to a different
graph-theoretic quantity. The associated problem would then closely resemble
scenarios where we aim to optimize some quantity over a matrix \emph{fitting}
a graph under specific constraints (see, \emph{e.g.}, the Lov\'{a}sz theta
function, the Colin de Verdi\`{e}re parameter, \emph{etc.}). This appears to
be an open research direction involving the relative entropy and possibly
other standard notions from quantum information theory.

\bigskip

\emph{Acknowledgments. }We would like to thank Koenraad Audenaert for
important discussion. VG acknowledges support from the Institut Mittag-Leffler
(Stockholm), where he was visiting while part of this work was done. SS is
supported by a Newton International Fellowship.


\begin{thebibliography}{99}                                                                                               %


\bibitem {he}K. M. R. Audenaert and J. Eisert, Continuity bounds on the
quantum relative entropy, \emph{J. Math. Phys.} \textbf{46}, 102104 (2005).

\bibitem {QHT}K. M. R. Audenaert, M. Nussbaum, A. Szkola, F. Verstraete,
Asymptotic Error Rates in Quantum Hypothesis Testing, \emph{Comm. Math. Phys.}
\textbf{279}, 251-283 (2008).

\bibitem {bgs}S. Braunstein, S. Ghosh, and S. Severini, The laplacian of a
graph as a density matrix: a basic combinatorial approach to separability of
mixed states, \emph{Ann. Comb.}, Vol. 10, no 3 (2006), 291-317.

\bibitem {bo}B. Bollob\'{a}s, \emph{Modern Graph Theory}, Springer Verlag, New
York, 1998.

\bibitem {ge}I. B. Gertsbakh, Y. Shpungin, \emph{Models of Network
Reliability: Analysis, Combinatorics, and Monte Carlo}, CRC Press, 2009.

\bibitem {hi}A. Hinchliffe (Editor), \emph{Chemical Modelling: Applications
and Theory, Volume 2}, The Royal Society of Chemistry (2002).

\bibitem {hol}A. S. Holevo, Bounds for the quantity of information transmitted
by a quantum communication channel, \emph{Prob. Inf. Transm.} (USSR)
\textbf{9}, 177--83 (1973).

\bibitem {ki}G. Kirchhoff, \"{U}ber die Aufl\"{o}sung der Gleichungen, auf
welche man bei der untersuchung der linearen verteilung galvanischer
Str\"{o}me gef\"{u}hrt wird, \emph{Ann. Phys. Chem.} \textbf{72}, 497-508, 1847.

\bibitem {och}W. Ochs, A new axiomatic characterization of the von Neumann
entropy, \emph{Rep. Math. Phys.} \textbf{8} (1975),109--120.

\bibitem {ni}A. Nielsen and I. L. Chuang, \emph{Quantum Computation and
Quantum Information}, Cambridge Series on Information and the Natural
Sciences, 2000.

\bibitem {na}N. Nakanishi, \emph{Graph Theory and Feynman Integrals}, Gordon
and Breach, New York, 1971.

\bibitem {pa}V. Paulsen, \emph{Completely Bounded Maps and Operator Algebras},
Cambridge University Press, Cambridge, 2002.

\bibitem {re}A. R\'{e}nyi, \emph{Probability Theory}, Amsterdam:
North-Holland, 1970.

\bibitem {sh}B. Schumacher and M. Westmoreland, Relative entropy in quantum
information theory.\ \emph{American Mathematical Society Contemporary
Mathematics Series: Quantum Information and Quantum Computation, 305},
American Mathematical Society, Providence, 2002.

\bibitem {st}R. P. Stanley, \emph{Enumerative combinatorics}, \emph{vol. I.},
Wadsworth and Brooks/Cole, Monterey, 1986.

\bibitem {tu}W. T. Tutte, \emph{Graph Theory}, Encyclopedia of Mathematics and
its Applications, 21, Addison-Wesley, 1984.

\bibitem {ve}V. Vedral, The Role of Relative Entropy in Quantum Information
Theory, \emph{Rev. Mod. Phys.} \textbf{74}, 197--234 (2002).

\bibitem {wat}W. Watkins, The Laplacian matrix of a graph: Unimodular
congruence, \emph{Linear and Multilinear Algebra} \textbf{28}:35-43 (1990).

\bibitem {we}D. Welsh, \emph{The Tutte polynomial. Random Structures
Algorithms}, \textbf{15(}3-4):210--228, 1999. Statistical physics methods in
discrete probability, combinatorics, and theoretical computer science
(Princeton, NJ, 1997).
\end{thebibliography}
\end{document}